# ON THE DYNAMICS OF NON-RIGID ASTEROID ROTATION

**Sergey V. Ershkov**

Affiliation: Plekhanov Russian University of Economics,

Scopus number 60030998, e-mail: sergej-ershkov@yandex.ru

**Dmytro Leshchenko**, Odessa State Academy of Civil Engineering

and Architecture, Odessa, Ukraine, e-mail: leshchenko_d@ukr.net

We have presented in this communication a new solving procedure for the dynamics of *non-rigid* asteroid rotation, considering the final spin state of rotation for a small celestial body (asteroid). The last condition means the ultimate absence of the applied external torques (including short-term effect from torques during collisions, long-term YORP effect, etc.).

Fundamental law of angular momentum conservation has been used for the aforementioned solving procedure. The system of *Euler* equations for dynamics of *non-rigid* asteroid rotation has been explored with regard to the existence of an analytic way of presentation of the approximated solution.

Despite of various perturbations (such as collisions, YORP effect) which destabilize the rotation of asteroid via deviating from the current spin state, the inelastic (mainly, tidal) dissipation reduces kinetic energy of asteroid. So, evolution of the spinning asteroid should be resulting by the rotation about maximal-inertia axis with the proper spin state corresponding to minimal energy with a fixed angular momentum.

Basing on the aforesaid assumption (component $K_1$ is supposed to be fluctuating near the given appropriate constant of the fixed angular momentum), we have obtained that 2-nd component $K_2$ is the solution of appropriate *Riccati* ordinary differential equation of 1-st order, whereas component $K_3$ should be determined via expression for $K_2$.

## 1. Introduction, the system of equations.

The main motivation of the current research is the analytical exploration of the dynamics of asteroid rotation when it moves in elliptic orbit through Space. In our previous research [1], we have explored the regimes of *rigid* asteroid rotation under the additional influence of YORP-effect. Let us note that assumption of asteroid rotating as rigid body means that distances between various points inside the rigid body should be preferably constant or should be elongated negligibly.

We will consider here regime of rotation of small celestial bodies (asteroids less than < 10 km in diameter) which differ from the rigid body in a sufficient extent. It means that distances between various points inside the asteroid can not be considered as being elongated negligibly. Meanwhile, only circa 20% of all the registered asteroids (near-Earth objects in NASA data base) are recognized as to be close to the rigid body approximation. For example, we can provide the comparison with to the actual/observed data for nonrigid asteroids, which is available in the modern research with respect to the *rubble pile asteroids* [2].

It is very important to create the adequate physical model along with the mathematical model of the aforementioned asteroid's spinning phenomenon with the main aim of the clarifying the results of data of astrometric observations. Indeed, if regime of rotation of asteroid is suddenly changing, we could observe even physical disintegration of asteroids (or self-destruction under the influence of sudden acceleration during a fast rotation [3]).

It is also worth to note that fundamental law of angular momentum conservation should be valid even during the *non-rigid* regime of asteroid's rotation [4]. Namely, theorem of conservation of angular momentum describes rotation of asteroid in a frame of reference fixed in the rotating body [5] ($I_i \neq 0$):

$$\frac{d\vec{K}}{dt} + [\vec{\Omega} \times \vec{K}] = \vec{M}, \qquad (1)$$

where $\vec{K} = \{ I_i \cdot \Omega_i \}$, whereas $\vec{\Omega} = \{ \Omega_i \}$ (here $\Omega_i$ are the components of angular velocity vector along the principal axes, i = 1, 2, 3), $I_i$ are the principal moments of inertia, and $\vec{M} = \vec{M}(t)$ is the total sum of applied external torques (including short-term effect from torques during collisions, long-term YORP effect [1], [6], etc.).

Let us especially empasize that we will consider here principal moments of inertia to be variable (time-dependent, $I_i = I_i(t)$), in general case; e.g., components of inertia tensor of asteroid may be changed during collisions [4]. Indeed, we should take into consideration the possible changes in its form, along with the decreasing of the mass via partial physical disintegration of asteroids during collisions or even via self-destruction due to the regime of fast rotation [1]. Despite of various perturbations (such as collisions, YORP effect) which destabilize the rotation of asteroid via deviating from the current spin state, the inelastic (mainly, tidal) dissipation [7-9] reduces kinetic energy of asteroid.

It means that evolution of the spinning asteroid should be resulting by the rotation about maximal-inertia axis [7] with the proper spin state corresponding to minimal energy with a fixed angular momentum.

We will consider in (1) only such the aforesaid final dynamical state of asteroid rotation (which is fluctuating near the given appropriate constant of the fixed angular momentum). Asteroid is supposed to be moving along its orbit far from the close influences of additional gravitational forces from planet of mass $m_{planet}$ or far from *Hill sphere* [1] (motion of asteroid is determined by equations of ER3BP with primaries $m_{planet}$ and $M_{Sun}$, $m_{planet} < M_{Sun}$):

$$r_H = a_p \cdot \left( \frac{m_{planet}}{3 M_{Sun}} \right)^{\frac{1}{3}} \qquad (*)$$

where $a_p$ is semimajor axis of the planet.

Let us also assume (as first approximation) that all external torques, associated with inertial forces, tides, YORP effect are neglected in (1) (i.e., $\vec{M} \cong \vec{0}$ in (1)).

According to the results of [7], inelastic (mainly tidal) dissipation, which is reducing kinetic energy, yields evolution of spin towards rotation about maximal-inertia axis $I_1$ with rate of rotation $\Omega_1$ (for definiteness, $I_1 > I_2 > I_3$); it means:

$$\{\Omega_2, \Omega_3\} \ll \Omega_1 \quad \Rightarrow \quad \{\Omega_2, \Omega_3\} \to 0 \qquad (2)$$

The last but not least, let us additionally note that the spatial ER3BP is not conservative, and no integrals of motion are known [10] (including total angular momentum, which combines the expressions in (1) and orbital angular momentum).

## 2. Analytical exploring of the system of equations (1).

First of all, we should note that (1) is the system of 3 nonlinear differential equations with respect to $\vec{K} = \{ I_i \cdot \Omega_i \}$ (with all coefficients depending on time *t*):

$$\frac{d\vec{K}}{dt} + [\vec{\Omega} \times \vec{K}] = \vec{0}, \quad \Rightarrow \quad \begin{cases} \dfrac{dK_1}{dt} = K_2 \cdot (\Omega_3) - K_3 \cdot (\Omega_2), \\[6pt] \dfrac{dK_2}{dt} = K_3 \cdot (\Omega_1) - K_1 \cdot (\Omega_3), \\[6pt] \dfrac{dK_3}{dt} = K_1 \cdot (\Omega_2) - K_2 \cdot (\Omega_1). \end{cases} \quad (3)$$

$$\Rightarrow \quad \begin{cases} \dfrac{dK_1}{dt} = K_2 \cdot (\dfrac{K_3}{I_3}) - K_3 \cdot (\dfrac{K_2}{I_2}), \\[6pt] \dfrac{dK_2}{dt} = K_3 \cdot (\dfrac{K_1}{I_1}) - K_1 \cdot (\dfrac{K_3}{I_3}), \\[6pt] \dfrac{dK_3}{dt} = K_1 \cdot (\dfrac{K_2}{I_2}) - K_2 \cdot (\dfrac{K_1}{I_1}). \end{cases} \quad \Rightarrow \quad \begin{cases} \dfrac{dK_1}{dt} = K_2 \cdot K_3 \cdot \left(\dfrac{I_2 - I_3}{I_2 \cdot I_3}\right), \\[6pt] \dfrac{dK_2}{dt} = K_1 \cdot K_3 \cdot \left(\dfrac{I_3 - I_1}{I_1 \cdot I_3}\right), \\[6pt] \dfrac{dK_3}{dt} = K_1 \cdot K_2 \cdot \left(\dfrac{I_1 - I_2}{I_1 \cdot I_2}\right). \end{cases} \quad (4)$$

Let us multiply 1-st equation of system (3) or (4) on $(K_1/I_1)$, the 2-nd Eq. on $(K_2/I_2)$, 3-rd on $(K_3/I_3)$; then if we sum all the resulting equations of the system above, we should obtain

$$\left(\frac{1}{I_1}\right)\frac{d(K_1^2)}{dt} + \left(\frac{1}{I_2}\right)\frac{d(K_2^2)}{dt} + \left(\frac{1}{I_3}\right)\frac{d(K_3^2)}{dt} = 0 \quad (5)$$

## 3. Solving procedure and the approximated solution for Eqns. (1).

According to the assumption (2) above, in (1) we will consider only the final dynamical state of asteroid rotation (which is fluctuating near the given appropriate constant of the fixed angular momentum, $K_1 \cong const$). It means that equation (5) can be transformed to the form below

$$\frac{d(K_2^2)}{dt} + \left(\frac{I_2}{I_3}\right)\frac{d(K_3^2)}{dt} \cong 0 \qquad (6)$$

Let us note that in case $K_1 \cong const$, 1-st equation of system (4) should be satisfied accordingly at first approximation (if we take into account assumption (2) for the right part of 1-st equation of system (4), where $\Omega_2 = (K_2/I_2)$, $\Omega_3 = (K_3/I_3)$, $\{\Omega_2, \Omega_3\} \to 0$).

As for the 2-nd equation of system (4), we obtain (here below $K_1 \cong const$):

$$K_3 = \frac{1}{K_1}\left(\frac{I_1 \cdot I_3}{I_3 - I_1}\right)\frac{dK_2}{dt} \qquad (7)$$

Now, as for the 3-rd equation of system (4), let us substitute expression for $K_3$ from the 2-nd Eqn. of (4) the expression for derivative in the left part; it yields as below

$$\left(\frac{I_1 \cdot I_3}{I_3 - I_1}\right)\cdot\frac{d^2 K_2}{dt^2} + \frac{d\left(\frac{I_1 \cdot I_3}{I_3 - I_1}\right)}{dt}\cdot\frac{dK_2}{dt} - \left(K_1^2 \cdot \left(\frac{I_1 - I_2}{I_1 \cdot I_2}\right)\right)\cdot K_2 = 0 \qquad (8)$$

where equation (8) for the dynamics of component $K_2 = I_2 \cdot \Omega_2$ could be transformed by change of variables $y = (K_2'/K_2)$ to the *Riccati* ODE of 1-st order [1].

**Discussion**

We have explored here the dynamics of *non-rigid* asteroid rotation, considering the final spin state of rotation for a small celestial body (asteroid). *Non-rigid* character of asteroid rotation means that principal moments of inertia are variable (time-dependent, $I_i = I_i(t)$, i = 1, 2, 3).

Fundamental law of angular momentum $\vec{K} = \{I_i \cdot \Omega_i\}$ conservation (which should be valid even during the *non-rigid* regime of asteroid's rotation) has been used at obtaining the analytical algorithm for solving. The proper approximate solution has been obtained which is presented below:

- component $K_1 = I_1(t) \cdot \Omega_1(t)$ is supposed to be fluctuating near the given appropriate constant of the fixed angular momentum, $K_1 \cong const$;

- component $K_2 = I_2(t) \cdot \Omega_2(t)$ is the solution of the appropriate *Riccati* ODE (8):

$$\left(\frac{I_1 \cdot I_3}{I_3 - I_1}\right) \cdot \frac{d^2 K_2}{dt^2} + \frac{d\left(\frac{I_1 \cdot I_3}{I_3 - I_1}\right)}{dt} \cdot \frac{dK_2}{dt} - \left(K_1^2 \cdot \left(\frac{I_1 - I_2}{I_1 \cdot I_2}\right)\right) \cdot K_2 = 0,$$

- component $K_3 = I_3(t) \cdot \Omega_3(t)$ is determined in (7) via expression for $K_2$:

$$K_3 = \frac{1}{K_1}\left(\frac{I_1 \cdot I_3}{I_3 - I_1}\right)\frac{dK_2}{dt}$$

We should additionally note that for reason of a special character of the solutions of *Riccati*-type ODEs, there exists a possibility for sudden *jumping* of magnitude of the solution at some meaning of time-parameter *t* [11-15].

Mathematical procedure of presenting the components of angular velocity via Euler angles [16] (and Wisdom angles [17]) has been demonstrated at the Appendix in [1].

Additional discussions regarding the current research see in Appendix **A2**.

## **Conclusion**

We have presented in this communication a new solving procedure for the dynamics of *non-rigid* asteroid rotation, considering the final spin state of rotation for a small celestial body (asteroid). The last condition means the ultimate absence of the applied external torques (including short-term effect from torques during collisions, long-term YORP effect, etc.).

Fundamental law of angular momentum conservation has been used for the aforementioned solving procedure. The system of *Euler* equations for dynamics of *non-rigid* asteroid rotation has been explored with regard to the existence of an analytic way of presentation of the approximated solution.

Despite of various perturbations (such as collisions, YORP effect) which destabilize the rotation of asteroid via deviating from the current spin state, the inelastic (mainly, tidal) dissipation reduces kinetic energy of asteroid. So, evolution of the spinning asteroid should be resulting by the rotation about maximal-inertia axis with the proper spin state corresponding to minimal energy with a fixed angular momentum.

Basing on the aforesaid assumption (component $K_1$ is supposed to be fluctuating near the given appropriate constant of the fixed angular momentum), we have obtained that 2-nd component $K_2$ is the solution of the appropriate *Riccati* ordinary differential equation of 1-st order, whereas component $K_3$ should be determined via expression for $K_2$.

There is additional condition for obtaining such approximated solution ($I_1 > I_2 \geq I_3$):

$$\{\Omega_2, \Omega_3\} << \Omega_1,$$

where $\Omega_i$ are the components of angular velocity vector along the principal axes (i = 1,2,3), $I_i$ are the principal moments of inertia.

The last but not least, we can obtain one additional class of approximated solutions of system (1) with non-zero external applied torques $\vec{M}(t) \neq \vec{0}$; mathematical procedure of obtaining such the additional solution has been moved to an Appendix **A1**, with only the resulting formulae left in the main text (here below $K_1 \cong const$):

$$K_2 \cdot \frac{dK_2}{dt} = M_2 \cdot K_2 + \left\{ M_1 \cdot K_1 \cdot \left( \frac{I_2}{I_3 - I_2} \right) \cdot \left( \frac{I_3 - I_1}{I_1} \right) \right\}, \qquad (10)$$

$$K_3 \cdot \frac{dK_3}{dt} = M_3 \cdot K_3 + \left\{ K_1 \cdot M_1 \cdot \left( \frac{I_3}{I_3 - I_2} \right) \cdot \left( \frac{I_1 - I_2}{I_1} \right) \right\}, \qquad (11)$$

where equations (10)-(11) for the dynamics of components of angular momentum $K_2$, $K_3$ ($K_2 = I_2 \cdot \Omega_2$, $K_3 = I_3 \cdot \Omega_3$) are the *Abel* ODEs of 1-st order of the 2-nd kind [13].

Also, the remarkable articles [18-20] should be cited, which concern the problem under consideration.

## **Acknowledgements**

## **Conflict of interest**

Authors declare that there is no conflict of interests regarding publication of article.

Remark regarding contributions of authors as below:
 In this research, Dr. Sergey Ershkov is responsible for the general ansatz and the solving procedure, simple algebra manipulations, calculations, results of the article in Sections 1-3 and also is responsible for the search of approximate solutions.
Dr. Dmytro Leshchenko is responsible for theoretical investigations as well as for the deep survey in literature on the problem under consideration (in Section 1, see the remark regarding ref. [2]).
Both authors agreed with the results and conclusions of each other in Sections 1-3.

**Appendix A1 (additional class of approximated solutions of system (1)).**

Let us obtain the additional class of approximated solutions of system (1). We consider the final dynamical state of asteroid rotation (which is fluctuating near the given appropriate constant of the fixed angular momentum, $K_1 \cong const$) for which we assume $\vec{M}(t) \neq \vec{0}$.

Then 1-st equation of system (1) should be satisfied accordingly (at first approximation) under the appropriate condition below:

$$K_2 \cdot K_3 \cdot \left( \frac{I_3 - I_2}{I_2 \cdot I_3} \right) \cong M_1 \qquad (9)$$

Meanwhile, there is no need to take into account assumption (2) for the right part of 1-st equation of system (1) in this case.

As for the 2-nd and 3-rd equations of system (1), we obtain (here below $K_1 \cong const$):

$$K_2 \cdot \frac{dK_2}{dt} = M_2 \cdot K_2 + \left\{ M_1 \cdot K_1 \cdot \left( \frac{I_2}{I_3 - I_2} \right) \cdot \left( \frac{I_3 - I_1}{I_1} \right) \right\}, \qquad (10)$$

$$K_3 \cdot \frac{dK_3}{dt} = M_3 \cdot K_3 + \left\{ K_1 \cdot M_1 \cdot \left( \frac{I_3}{I_3 - I_2} \right) \cdot \left( \frac{I_1 - I_2}{I_1} \right) \right\}, \qquad (11)$$

where equations (10)-(11) for the dynamics of components of angular momentum $K_2$, $K_3$ ($K_2 = I_2 \cdot \Omega_2$, $K_3 = I_3 \cdot \Omega_3$) are the *Abel* ODEs of 1-st order of the 2-nd kind. These Eqns. can be transformed by the appropriate change of variables $K_3 = 1/u$ to the *Abel* ODEs of the 1-st kind (of *Riccati* type).

Accordingly, for the aforesaid reason of a special character of the solutions of *Riccati*-type ODEs (see **Discussion**), there exists a possibility for sudden *jumping* of magnitude of the solution at definite meaning of time-parameter $t$.

In the physical sense, such jumping of *Riccati*-type solutions of Eqn. (8) can be associated with the effect of sudden acceleration/deceleration of angular velocity's component $\Omega_2$ at definite moment of time $t_0$ (or with the alternative effect of crucial changes in the principal moment of inertia $I_2$ ($t$) of asteroid during the process of rotation).

___________________________________________________________

# Appendix A2 (additional discussions regarding the presented research).

# Reply on the Comments on

# "ON THE DYNAMICS OF NON-RIGID ASTEROID ROTATION"

## Abstract

We have presented here our reply on the comments on our recent article "ON THE DYNAMICS OF NON-RIGID ASTEROID ROTATION".

Against our partial solution stems from fundamental law of angular momentum conservation for the non-rigid asteroid rotation (where asteroids were considered preferably deforming as *'rubble pile' asteroids*, not as elastic bodies), authors of the aforementioned comments have considered other partial solving procedure which can be associated with commensurable elongations of the asteroid during the deformations. This assumption contradicts the physical nature most of known *asteroids* (*'rubble pile' asteroids* are known to have a non-elastic internal rheology), so such the solutions of *Euler* equations for dynamics of *non-rigid* asteroid rotation, obtained by authors of the comments, have no real astrophysical sense.

Despite of useless in astrophysical area, the suggested strategy or solving procedure for the constructing a new class of special kind of solutions for *Euler* equations (dynamics of *non-rigid* elastic body's rotation) can be of partial interest to a journals for theoretical mechanics but not first-class astrophysical journals or journal in the field of celestial mechanics. So, we wish for the authors every success to possible publishing their full solution (not a simple a note, which seems to be "following-in-wake" in regard to our article) where they could discuss additionally the stability of solutions if required.

1. **Introduction, the system of equations.**

First, we should appreciate the efforts of Scientific Editor of "Acta Astronautica" who is paying additional attention to our recently published article [1] which developing the ideas previously published in [2]. Indeed, every additional discussion regarding the published article should increase its impact as well improve the understanding of the main ideas of the suggested researches.

In [3], authors of such the comments have considered the alternative partial solving procedure (alternative with respect to partial solving procedure which has been suggested in [1] earlier) during which, for unkown reason, they came to conclusion of mistake in our solving procedure. Let us clarify their misleading in the key points which apparently differ not only from our (also, the partial case of solution) but their result is physically senseless solution.

Let us remind them that in [1] we have considered (see Introduction section) regime of rotation of small celestial bodies (asteroids less than < 10 km in diameter) which differ from the rigid body in a sufficient extent. Meanwhile, we stated that only circa 20% of all the registered asteroids are recognized as to be close to the rigid body approximation. So, we have constructed our solving procedure for the non-rigid asteroids, known as *'rubble pile' asteroids* [3-4]. Our having paid a special attention in [1] (which, nevertheless, was missed by the authors of comments [3]), that it is very important to create the adequate physical model along with the mathematical model of the aforementioned asteroid's spinning phenomenon. Namely, we have especially emphasized that we would consider in [1] all the principal moments of inertia to be variable (time-dependent, $I_i = I_i(t)$), in general case.

This assumption should not mean (as suggested by authors of comments [3]) that asteroid, being *rubble pile*, is commensurable elongated in all directions during the deformations. Obviously, the inelastic rheology of asteroid first yields elongations along the minimal-inertia axes $\{I_2, I_3\}$ instead of elongations along the maximal-inertia axis with rate of rotation $\Omega_1$ (we have chosen in [1]: $I_1 \gg I_2 > I_3$).

Nevertheless, thanks by occasion we should point out (or claim the correction) in regard to technical typo in [1-2]: namely, the *Hill sphere* (motion of asteroid is determined by equations of ER3BP with primaries $m_{planet}$ and $M_{Sun}$, $m_{planet} < M_{Sun}$) should be defined in [1-2] as below:

$$r_H = a_p \cdot \left(\frac{m_{planet}}{3 M_{Sun}}\right)^{\frac{1}{3}} \qquad (*)$$

where $a_p$ is semimajor axis of the planet.

The last but not least: authors of [3] claim that despite the challenging problem definition found in the introduction and highlights of [1], a simplified approximation is covered. As far as we have shown above, despite of a loud contraversion, they simply have introduced their own partial way of solving equation (5) (Eqn. (5) in our article [1]) without physically understanding of the suggested solutions (which can not be applied to the dynamics of *'rubble pile'* asteroid rotation).

Finally, authors of [3] claim in the Introduction that "An elliptic orbit, stated in the Highlights, may substantially complicate the dimensionless equations of motion that utilize the derivative with respect to the argument of latitude instead of time. As the dimensionless equations are not presented in the paper, this phenomenon is not actually covered". Here below is our Highlights (printed in [1]):

**Highlights**

- The dynamics of asteroid rotation moving along the elliptic orbit is explored.
- New type of solutions for dynamics of non-rigid asteroid rotation are obtained.
- New method for solving Euler's equations of non-rigid body rotation is suggested.
- Evolution of spin towards rotation about maximal-inertia axis is approximated.
- The 2-nd component of angular momentum is proved to be solution of Riccati ODE.

We wonder which of them authors of [3] are discussing above? It is simply a false.

2. **Discussion of equations (5) and (6).**

First of all, we should note that we suggested in [1] the differential invariant as below

$$\left(\frac{1}{I_1}\right)\frac{d(K_1^2)}{dt} + \left(\frac{1}{I_2}\right)\frac{d(K_2^2)}{dt} + \left(\frac{1}{I_3}\right)\frac{d(K_3^2)}{dt} = 0 \qquad (5)$$

where $\vec{K} = \{I_i \cdot \Omega_i\}$, whereas $\vec{\Omega} = \{\Omega_i\}$ (here $\Omega_i$ are the components of angular velocity vector along the principal axes, i = 1, 2, 3), $I_i$ are the principal moments of inertia.

In [3], authors claim that the ultimate result of the paper [1] that is the equations for the angular momentum components perpendicular to the maximal-inertia axis $I_1$ should be correct only under specific assumptions. Then they analyzed small deviations from the rotation around the maximum inertia axis, where component of the angular momentum (associated with the maximal-inertia axis $I_1$) is not neglected. As they obtained, the omitted terms in the equations of motions have the same, or even higher, order of magnitude as the retained terms. But such result is contradicted to the physical nature of asteroid of *'rubble pile'*-type, so authors of [3] have simply considered very special solutions of *Euler* equations for dynamics of *non-rigid* asteroid rotation which have no astrophysical sense. Indeed, acording to the data of modern astrophysical observations (in the 'near-Earth objects' NASA data base; see also [5]) only circa 20% of all the registered asteroids are recognized as to be close to the rigid body approximation [6], other 80% are recognized as to be the *'rubble pile'* asteroids. Among them, there are no asteroids with elastic rheology! (as e.g. ideal rubber).

It is more than unclear when authors claim that 'in [1] no assumption on the nature of the moments of inertia change is made' (see discussion regarding the *'rubble pile'* asteroids here above and in [1]).

Then authors of [3] claim their own assumption (without references, appropriate validation or justification) that "relation (2) in [1] may be applied to the angular momentum under proper and realistic assumptions for the asteroid" (as for corresponding stability analysis, we suggest them publishing their full solution where they could discuss additionally the stability of solutions if required). Immediately after, they suggested such the "proper and realistic assumptions" which lead to their own way of solving our equation (5), instead of our (6). So, authors of [3] have suggested their own partial solving procedure, during which they have come to a wrong conclusion regarding our way of solving (5).

Here we should stop the discussion because our solving procedure is based on realistic astrophysical nature of the *'rubble pile'* asteroids, their partial solving procedure [3] (based on analysis of their equation (3)) is not. Their equation (3) stems from the wrong assumption that asteroid should be commensurably elongated in all directions during the deformations. But, alas, this is not so for the *'rubble pile'* asteroids as explained above. This is the main mistake of authors of comments in [3].

So, all other comments in [3] are false if we discarded their Eqn. (3) due to astrophysical senseless.

Despite of useless in astrophysical area, the suggested strategy or solving procedure for the constructing a new class of special kind of solutions for *Euler* equations (dynamics of *non-rigid* elastic body's rotation) can be of partial interest to a journals for theoretical mechanics but not first-class astrophysical journals or journal in the field of celestial mechanics. The only true comment of authors of [3] is concerning the "some new variable $K_2$" during transformation of equation of Riccati-type (8). Indeed, it was the type-set typo (Editors of the journal can verify this fact): we have approved our final version of the article where it was typeset as $K_2'$ (but in the published version it has been turned for unknown reasons as $K_{2'}$). So, thanks to opportunity to point out here: we should note that it should be denoted as $K_2' = dK_2/dt$.

After that, authors of [3] have claimed that "equation (8) is not solved". Indeed, equation of Riccati-type has no general solution and this is not our fault [7].

**Conclusion**

Overall, the proposed criticism in [3] is based on unphysical assumption, which leads authors of [3] to erroneous conclusion. They should take into consideration the special literature on celestial mechanics regarding the problem under consideration [4-6] and also we recommend to search other researches by Dr. Walsh (on the theme of *'rubble pile'* asteroids structure, rheology and dynamics).

Other discussion regarding discarding the suggested in [1] "new method for solving Euler equations" (and other results of [1]) is dubious.

The comments [3] is almost void except pointing out the typo (as noted above) regarding "some new variable $K_{2'}$"; namely, [3] contains the attempts to suggest their own partial solution of our Eqn. (5), but based on the unrealistic assumption for the regimes of rotations for the *'rubble pile'* asteroids.

At last, authors of [3] claim that Appendix in [1] is not related to the paper. This is simply an example of invalid logic, because this part of [1] is directly linked to the main body of the paper by the mathematical procedure of obtaining the additional solution which has been moved to an Appendix.

The main equation (3) presented in [3] is proved to be non-adequate to the physical model of the *'rubble pile'* asteroids rotation. All in all, the structure, aim and discussion in the comments [3] are found to be based on unreasonable assumptions (such the comments should not be paying attention by specialists in celestial mechanics who works in the discussed theme ('*rubble pile*' asteroid dynamics) and related areas).